\documentclass[twocolumn,tighten]{aastex62}
\usepackage{isotope}
\usepackage{breakurl}

\received{}
\revised{}
\accepted{}
\submitjournal{ApJL}

\shorttitle{\isotope[3]{He}-Rich Solar Energetic Particles}
\shortauthors{Bu\v{c}\'ik et al.}

\begin{document}

\title{\isotope[3]{He}-Rich Solar Energetic Particles from Sunspot Jets}

\correspondingauthor{Radoslav Bu\v{c}\'ik}
\email{bucik@mps.mpg.de}


\author{Radoslav Bu\v{c}\'ik}
\affiliation{Institut f\"{u}r Astrophysik, Georg-August-Universit\"{a}t G\"{o}ttingen, D-37077 G\"{o}ttingen, Germany}
\affiliation{Max-Planck-Institut f\"{u}r Sonnensystemforschung, D-37077 G\"{o}ttingen, Germany}

\author{Mark E. Wiedenbeck}
\affiliation{Jet Propulsion Laboratory, California Institute of Technology, Pasadena, CA 91109, USA}

\author{Glenn M. Mason}
\affiliation{Applied Physics Laboratory, Johns Hopkins University, Laurel, MD 20723, USA}

\author{Ra\'ul G\'omez-Herrero}
\affiliation{Space Research Group, University of Alcal\'a, E-28871 Alcal\'a de Henares, Spain}

\author{Nariaki V. Nitta}
\affiliation{Lockheed Martin Advanced Technology Center, Palo Alto, CA 94304, USA}

\author{Linghua Wang}
\affiliation{Institute of Space Physics and Applied Technology, Peking University, 100871 Beijing, China}



\begin{abstract}

Solar sources of suprathermal ($<$1\,MeV\,nucleon$^{-1}$) $^3$He-rich solar energetic particles (SEPs) have been commonly associated with jets originating in small, compact active regions at the periphery of near-equatorial coronal holes. Sources of relatively rare, high-energy ($>$10\,MeV\,nucleon$^{-1}$) $^3$He-rich SEPs remain unexplored. Here we present two of the most intense $^3$He-rich ($^3$He/$^4$He$>$1) SEP events of the current solar cycle 24 measured on the {\sl Advanced Composition Explorer} at energy $>$10\,MeV\,nucleon$^{-1}$. Although $^3$He shows high intensities, $Z$$>$2 ions are below the detection threshold. The events are accompanied by type-III radio bursts, but no type-II emission as typically seen for suprathermal $^3$He-rich SEPs. The corresponding solar sources were analyzed using high-resolution, extreme-ultraviolet imaging and photospheric magnetic field observations on the {\sl Solar Dynamics Observatory}. We find the sources of these events associated with jets originating at the boundary of large sunspots with complex $\beta\gamma\delta$ magnetic configuration. Thus, details of the underlying photospheric field apparently are important to produce $^3$He to high energies in the examined events. 

\end{abstract}



\section{Introduction} \label{sec:intro}

$^3$He-rich solar energetic particles (SEPs) are characterized by a peculiar ion composition markedly different from the corona or solar wind \citep[e.g.,][]{koc84,mas07}. The abundance of $^3$He is enhanced by factors up to 10$^4$; heavy (Ne--Fe) and ultra-heavy ions ($Z$$>$30) show enhancement by factors $\sim$3--10 and $>$100, respectively, independently of $^3$He enhancement \citep{mas86,mas04,rea94,rea04}. $^3$He-rich SEPs are firmly associated with type-III radio bursts \citep[e.g.,][]{nit06} and their parent low-energy electrons \citep[e.g.,][]{wan12}. A distinct composition of $^3$He-rich SEPs is believed to indicate a unique acceleration mechanism operating in their solar sources. Variety of models have been proposed for ion acceleration in $^3$He-rich SEP events \citep[see review by][]{mil98}. Most models involve ion-cyclotron resonance with plasma waves.

Solar sources of $^3$He-rich SEPs have been associated with coronal jets \citep[e.g.,][]{kah01,wan06a,nit06,nit08,nit15,buc14,buc18,che15}, indicating acceleration in magnetic reconnection on field lines open to interplanetary space \citep[e.g.,][]{shi00}. The events with high $^3$He and heavy-ion enrichments have shown unwinding motions in jets \citep{mas16,inn16,buc18}. In some events, jets were accompanied by large-scale propagating coronal fronts \citep{nit15,buc16}. Jets in $^3$He-rich SEP events have been found to originate at the compact, small active regions near the coronal holes \citep{wan06a,pic06,buc18}. There are reports on $^3$He-rich jets from a plage region \citep{che15} or sunspot umbra \citep{nit08}. All these studies are based on ion measurements below a few MeV\,nucleon$^{-1}$. Solar sources of relatively rare, high-energy ($>$10\,MeV\,nucleon$^{-1}$) $^3$He-rich SEP events remain unexplored although these events were for the first time detected at 10--100\,MeV\,nucleon$^{-1}$\citep{hsi70}. 

In this Letter, we have examined the solar sources of the two most intense high-energy ($>$10\,MeV\,nucleon$^{-1}$) $^3$He-rich SEP events of the present solar cycle 24 (2008--2017). We report that solar sources of these events are associated with jets originating at the boundary of large sunspots with complex magnetic configuration.

\section{Methods} \label{sec:method}

Reported in this paper, two $^3$He-rich SEP events on 2011 February 18 and 2015 August 24, were identified using measurements from the Solar Isotope Spectrometer \citep[SIS;][]{sto98} on the {\sl Advanced Composition Explorer} ({\sl ACE}). The SIS is a $dE/dx$ versus residual energy telescope, measuring He to Zn nuclei from $\sim$4 to $\sim$100\,MeV\,nucleon$^{-1}$. The reported events show the highest $^3$He intensities among all other highly enriched ($^3$He/$^4$He$>$1) SEP events at energy 10.5\,MeV\,nucleon$^{-1}$ in the examined ten-year period. Note that in the solar wind $^3$He/$^4$He$\sim$4$\times$10$^{-4}$ \citep[e.g.,][]{glo98}. The EPHIN \citep{mul95} measurements aboard SOHO confirm that these two events show the highest $^3$He intensities in the energy range 5--25\,MeV\,nucleon$^{-1}$. The EPHIN has a similar measurement principle as SIS but a smaller geometric factor. Low-energy abundances were examined using the Ultra Low Energy Isotope Spectrometer \citep[ULEIS;][]{mas98} on {\sl ACE}. The ULEIS is a time-of-flight mass spectrometer that measures ions in the energy range from 20\,keV\,nucleon$^{-1}$ to several MeV\,nucleon$^{-1}$. We also make use of energetic electron measurements made by 3DP EESA-H and SST telescopes \citep{lin95} on Wind, solar wind measurements made by SWEPAM \citep{mcc98} and MAG \citep{smi98} on {\sl ACE}. {\sl ACE}, SOHO, and Wind were in the interplanetary space at the L1 point.

The solar sources of \isotope[3]{He}-rich SEPs were investigated using high-resolution observations from the Atmospheric Imaging Assembly \citep[AIA;][]{lem12} and Helioseismic and Magnetic Imager \citep[HMI;][]{sch12} on the {\sl Solar Dynamics Observatory} ({\sl SDO}). The AIA provides full-disk images of the Sun with 1.5$\arcsec$ spatial and 12 s temporal resolution at seven EUV and three UV wavelength channels. We use 304, 171 and 193\,{\AA} channels that observe emissions from \ion{He}{2} ($\sim$0.08\,MK), \ion{Fe}{9} ($\sim$0.8\,MK) and \ion{Fe}{12} ($\sim$1.5\,MK) lines, respectively. The HMI provides full-disk line-of-sight magnetograms and continuum intensity images with a spatial resolution of 1$\arcsec$ and a cadence of 45\,s. We also inspect Wind WAVES \citep{bou95} radio spectra for the event associated type-III burst. The frequency range of WAVES ($<$14\,MHz) covers emissions generated from about two solar radii to 1\,au. Furthermore, we make use of soft X-ray observations from the NOAA GOES-15 XRS sensor.

\section{Results} \label{sec:res}

\subsection{\isotope[3]{He}-rich SEP events} \label{subsec:ev}

\begin{figure*}
\epsscale{1.17}
\plotone{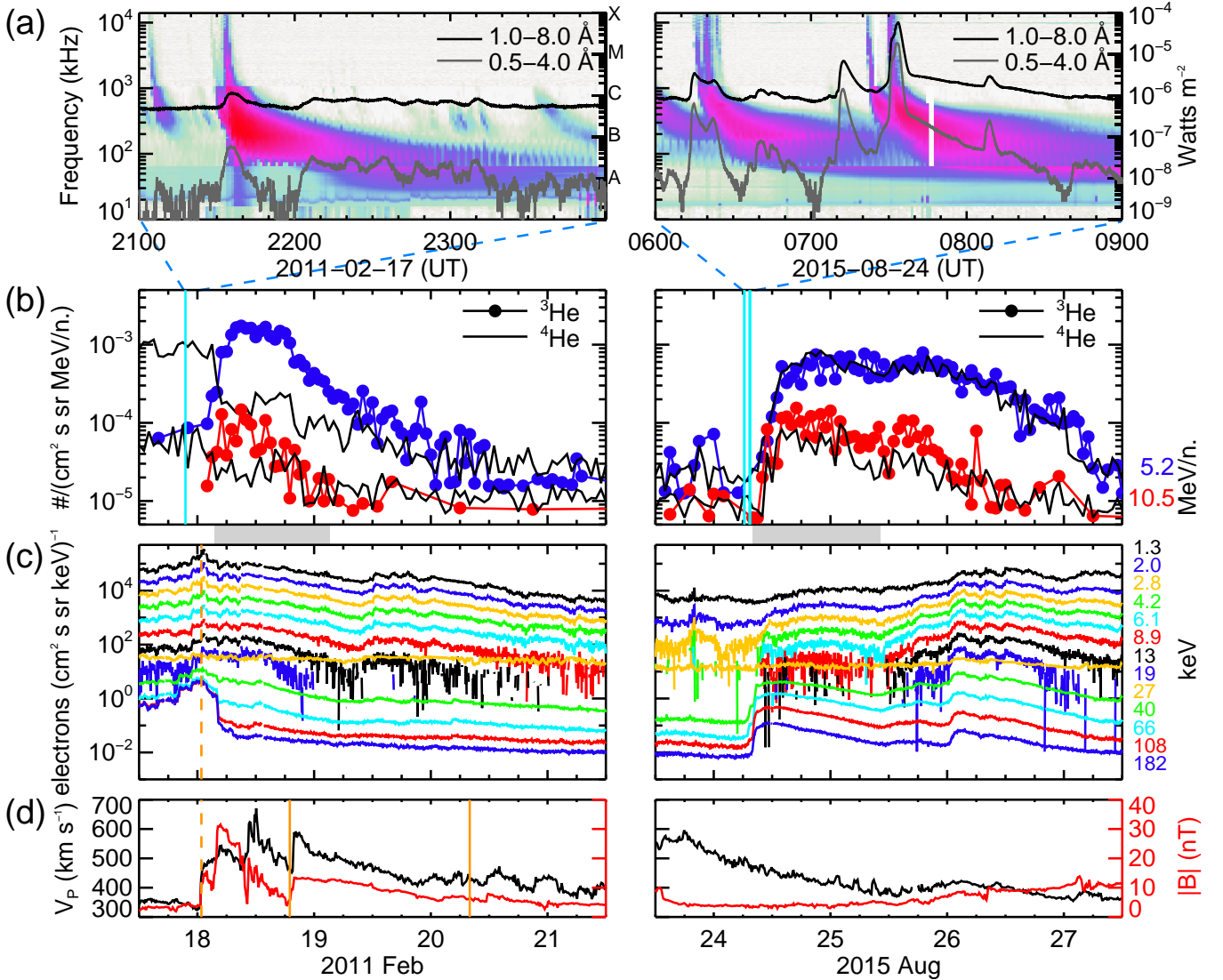}
\caption{(a) 1 minute Wind/WAVES radio spectrogram and GOES-15 2\,s X-ray fluxes (\deleted{black}two curves). The labels A, B, C, M, X indicate flare classes in the 1--8\,{\AA} channel. (b) 1 hr {\sl ACE}/SIS 5.2 and 10.5\,MeV\,nucleon$^{-1}$ \isotope[3]{He}, \isotope[4]{He} intensities. The upper (lower) curve shows the lower (higher)-energy \isotope[4]{He} intensities. Solid vertical lines mark the event-associated type-III bursts shown in panel (a). Grey shaded bars mark the periods considered for analysis in each event.\deleted{The dashed lines between the panels (a) and (b) indicate the time ranges of the upper panels with respect to the ion time profiles.} (c) 5 minute electron intensities from Wind/3DP (1.3--182\,keV). The dashed vertical line marks the IP shock. (d) 10 minute solar wind speed $V_p$ (black) and magnetic field magnitude $|B|$ (red). Solid vertical lines mark the ICME time interval.
 \label{fig:f1}}
\end{figure*}

Figure \ref{fig:f1} shows the energetic particle and solar wind plasma measurements for two examined \isotope[3]{He}-rich SEP events and the associated radio and soft X-ray activity. Figure \ref{fig:f1}(a) displays Wind/WAVES radio spectrograms and the GOES\deleted{1--8\,{\AA}} X-ray fluxes. Figure \ref{fig:f1}(b) shows 1 hr {\sl ACE}/SIS \isotope[3]{He}, \isotope[4]{He} intensities at 5.2 and 10.5\,MeV\,nucleon$^{-1}$. \isotope[4]{He} intensities below $\sim$10$^{-5}$ should be regarded as upper limits. Figure \ref{fig:f1}(c) presents Wind/3DP electron intensities at different energy channels (EESA-H 1.3--19\,keV and SST 27--182\,keV). Figure \ref{fig:f1}(d) shows the solar wind speed and magnetic field magnitude. 

\begin{deluxetable*}{lccccchcccchcch}
\tabletypesize{\scriptsize}
\tablewidth{0pt} 
\tablenum{1}
\tablecaption{$^3$He-rich SEP event properties \label{tab:tab1}}
\tablehead{\\
\colhead{SEP} & \colhead{Days\tablenotemark{a}} & \colhead{\isotope[3]{He}/\isotope[4]{He}\tablenotemark{a}} & \colhead{\isotope[3]{He} Flux\tablenotemark{a}} & \colhead{\isotope[3]{He}/\isotope[4]{He}\tablenotemark{b}} & \colhead{Fe/O\tablenotemark{b}} & \nocolhead{El.\tablenotemark{c}} & \colhead{Type-III\tablenotemark{c}}  & \multicolumn{4}{c}{Flare\tablenotemark{d}} & \multicolumn{2}{c}{CME\tablenotemark{e}} & \nocolhead{L1 Footpoint}\\
\cline{9-12}
\cline{13-14}
\colhead{Start} & \colhead{} & \colhead{} & \colhead{($\times10^{-5}$)} &  \colhead{} &  \colhead{} &  \nocolhead{Event} & \colhead{Start} &  \colhead{Class} &\colhead{Start} & \colhead{Loc.} & \nocolhead{NOAA AR} & \colhead{Speed} & \colhead{Width} & \nocolhead{Longitude}
}
\startdata
2011 Feb 18 & 049.17--050.13 & 2.33$^{+0.22}_{-0.17}$ & 5.26$^{+0.49} _{-0.38}$ & 0.12$\pm$0.01 & 1.46$\pm$0.13  & \nodata & Feb 17 21:32 & C1.1 & 21:30 & S19W46 & 11158  & 490 & 82 & W66 \\
2015 Aug 24 & 236.33--237.42 & 1.61$^{+0.09}_{-0.07}$ & 5.78$^{+0.33} _{-0.27}$ & 0.24$\pm$0.11 & 1.16$\pm$0.74 & 40--108 & Aug 24 06:16 & \nodata & \nodata & S14E01 & 12403 & \nodata & \nodata & W46\\
{} & {} & {} & {} & {} & {} & {} & Aug 24 06:20 & C2.3 & 06:20 & {"} & " & \nodata & \nodata & " \\
{} & {} & {} & {} & {} & {} & 2--310 & Aug 24 07:22 & C1.3 & 07:22 & S14W00 & {"} & \nodata & \nodata & W49\\
{} & {} & {} & {} & {} & {} & {} & Aug 24 07:30 & M5.6 & 07:26 & {"} & {"} & 272 & 88 & W50\\
\enddata
\tablenotetext{a}{10.5\,MeV\,nucleon$^{-1}$; flux units -- particles (cm$^2$\,s\,sr\,MeV/nuc.)$^{-1}$}
\tablenotetext{b}{0.546\,MeV\,nucleon$^{-1}$}
\tablenotetext{c}{$\sim$10\,MHz from Wind WAVES 1-minute data}
\tablenotetext{d}{GOES X-ray Class \& Start; Location in {\sl SDO}/AIA 304\,{\AA}}
\tablenotetext{e}{Speed (km\,s$^{-1}$), width ($^{\circ}$) from SOHO/LASCO catalog (\url{http://cdaw.gsfc.nasa.gov/CME_list})}
\end{deluxetable*}

The 2011 February 18 event occurred during a decay phase of the gradual SEP event of 2011 February 15 and near the passage of an interplanetary (IP) fast forward shock on 2011 February 18 00:49\,UT. Figure \ref{fig:f1} indicates that after the shock passage the \isotope[3]{He} intensity starts to increase while \isotope[4]{He} intensity decreases. The opposite time-intensity profiles suggest that the event onset near the shock passage was most likely coincidental. The shock on Wind has been reported at Harvard-Smithsonian Center for Astrophysics IP Shock Database (\url{https://www.cfa.harvard.edu/shocks/}). The corresponding IP coronal mass ejection (ICME) can be found in the list of the Near-Earth ICMEs (\url{http://www.srl.caltech.edu/ACE/ASC/DATA/level3/icmetable2.html}). To confirm the solar origin of the 2011 February 18 event we examined the ULEIS/SIS He mass spectrograms of individual ions in the energy ranges 0.4--1.0 and 7.6--16.3\,MeV\,nucleon$^{-1}$, available at the {\sl ACE} science center (\url{http://www.srl.caltech.edu/ACE/ASC/DATA/level3/sis/heplots}). The spectrograms indicate the velocity dispersive onset for the 2011 February 18 event. In 0.4--1.0 and 7.6--16.3\,MeV\,nucleon$^{-1}$ energy ranges the first \isotope[3]{He} ions appeared at $\sim$05--06\,UT and $\sim$02--03\,UT, respectively. The time lag between the onsets (3\,hr) is consistent with the propagation time difference ($257-63$ minutes) for ions in these two energy intervals traveling from Sun to L1 along the nominal Parker field line. The 2011 February 18 event is included in a statistical study of Fe-rich SEP events in 1995--2013 \citep{rea14} with Fe/O~$=1.34\pm0.20$ (at 3--4\,MeV\,nucleon$^{-1}$). Note that the intensity of $Z$$>$2 ions, including Fe, at energy $>$10\,MeV\,nucleon$^{-1}$ on SIS was below the measurable level for both events. The ULEIS confirms that at low energies both events are Fe-rich with Fe/O$\sim$1 (see Table \ref{tab:tab1}). Soft energy spectra and low $Z$$>$2 abundances (relative to He) in $^3$He-rich events frequently preclude SIS detection of these elements.

\begin{figure}
\epsscale{1.16}
\plotone{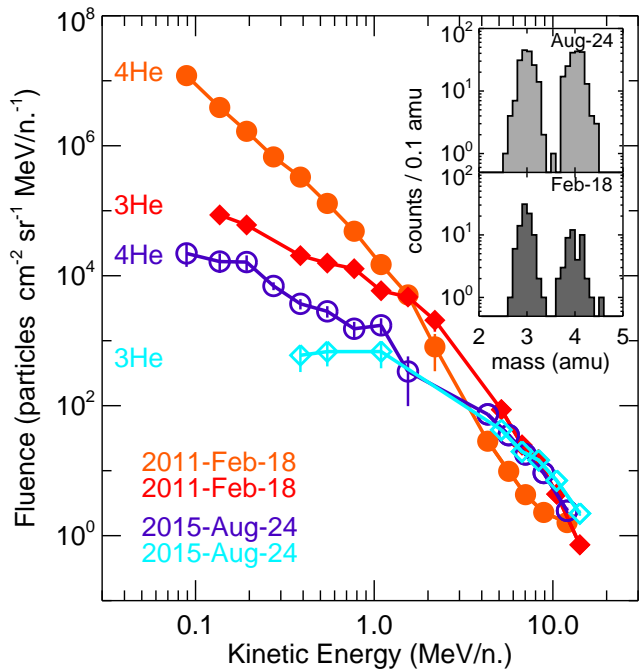}
\caption{Energy spectra of \isotope[4]{He} (circles) and \isotope[3]{He} (diamonds) for the 2011 February 18 (solid) and 2015 August 24 (open symbols) events measured with ULEIS (eight \isotope[3]{He} and ten \isotope[4]{He} energy bins; see text for an explanation of the missing spectral points) and SIS (five energy bins). Inset: \isotope[3]{He} (\isotope[4]{He}) mass histograms at 10.5 (8.9)\,MeV\,nucleon$^{-1}$. The SIS energy intervals correspond to intervals of \replaced{ion range}{ranges of ions (depths in the detector stack)}, resulting in a slightly lower energy per nucleon for \isotope[4]{He} than for \isotope[3]{He}. \label{fig:f2}}
\end{figure}

Solar energetic electrons were not clearly measured in the 2011 February 18 event due to enhanced background related to the preceding gradual SEP event and the passage of the IP shock. Furthermore, the electron channels at $\geq$108\,keV were contaminated by protons. Note that the SST data were corrected for scattered electrons that leave only a fraction of their energy in the detector \citep{wan06b}. An association with type-III radio burst is rather straightforward as there were no other significant type-III bursts in the 11-hr interval before the event start time. The event associated type-III burst on February 17 21:32\,UT was accompanied by a C3.1 X-ray flare (see Figure \ref{fig:f1}(a), left panel). The associations are consistent with the analysis by \citet{rea14}. The 2015 August 24 event was accompanied by two consecutive solar energetic electron events, separated by $\sim$1\,hr. For this reason, the first electron intensity enhancement, notably weaker (measured at $<100$\,keV) than the second, is not well resolved. Each electron event is associated with a double type-III burst; in the first electron event, the type-III bursts are separated by 4 minutes and in the second event by 8 minutes. The type-III bursts on August 24 06:16 and 06:20\,UT, related to the first electron event, occurred near a double X-ray flux enhancement (see Figure \ref{fig:f1}(a), right panel) corresponding to the C3.2 and C2.3 flares. The high cadence EUV images suggest that type-III burst at 06:16\,UT is likely unrelated to the C3.2 flare (see Section \ref{subsec:ss}). The type-III bursts on August 24 7:22 and 7:30\,UT, related to the second electron event, were accompanied by C1.3 and M5.6 flares, respectively. The C1.3 flare, not well resolved from elevated X-ray flux during the decay phase of previous stronger flare, was identified with the help of 0.5--4\,{\AA} GOES channel. None of the examined events have type-II bursts arising from coronal shocks. The events were accompanied by only slow CMEs (see Table \ref{tab:tab1}).

Figure \ref{fig:f2} shows \isotope[4]{He}, \isotope[3]{He} ULEIS and SIS fluence spectra for the two examined events. The He mass histograms integrated over the duration of the events are embedded in Figure \ref{fig:f2}. The \isotope[4]{He} fluence in the February 18 event should be treated as upper limit due to the possibility of a contribution from the preceding event. Note that the \isotope[3]{He} intensity increase on August 25 (see Figure \ref{fig:f1}(b), right panel), probably associated with the same active region as the August 24 event, is not included in the calculation of the event characteristics. Missing ULEIS spectral points (five for \isotope[3]{He} and one for \isotope[4]{He}) for the August 24 event means that the fluence was near the lowest measurable level (zero or one count in the integrated period). It is surprising that below $\sim$1\,MeV\,nucleon$^{-1}$ the 2015 August 24 event is at the threshold level, though at $>$10\,MeV\,nucleon$^{-1}$ it is one of the most intense \isotope[3]{He}-rich events in the solar cycle. The energy spectra are roughly consistent with double power laws where slopes are harder below $\sim$1.5\,MeV\,nucleon$^{-1}$. \deleted{In both events, the \isotope[3]{He} spectra are distinctly harder than \isotope[4]{He} below $\sim$1.5\,MeV\,nucleon$^{-1}$ causing a decrease of the \isotope[3]{He}/\isotope[4]{He} ratio towards low energies.}

\begin{figure*}
\epsscale{1.16}
\plotone{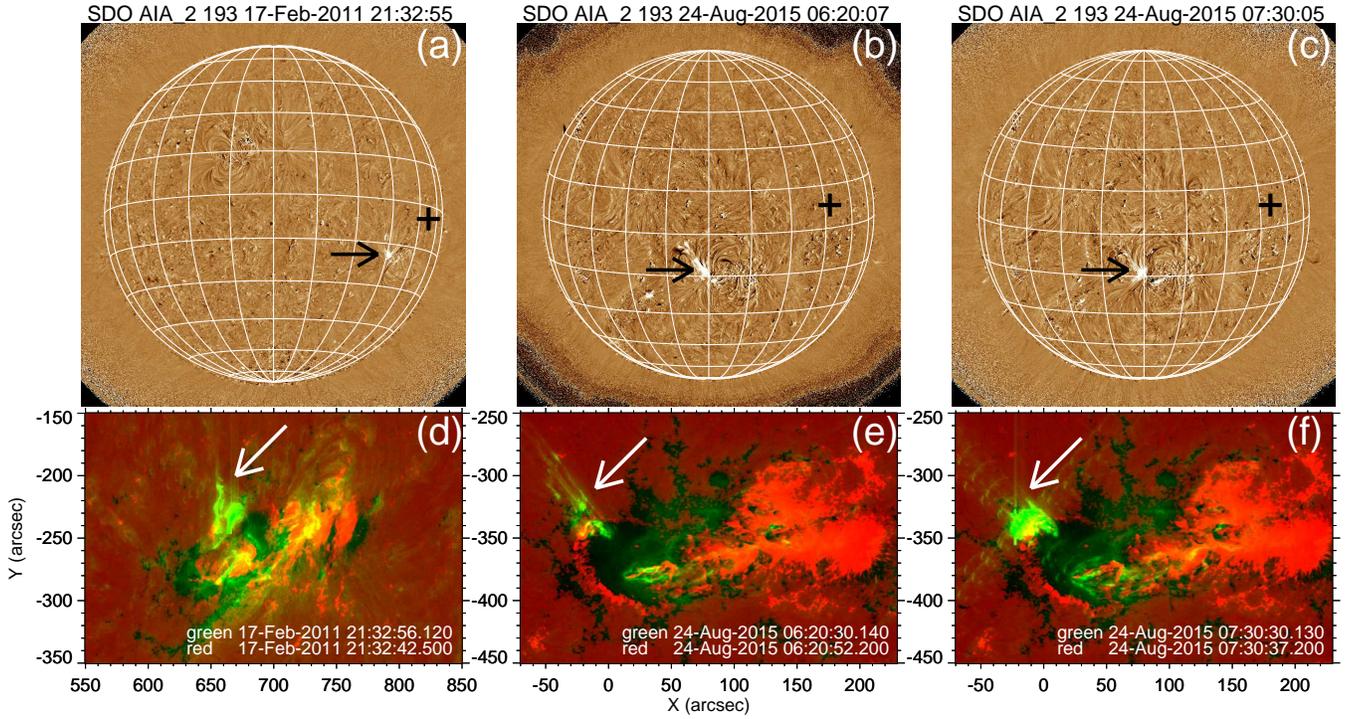}
\caption{(a)--(c) {\sl SDO} AIA 193\,{\AA} 1 minute base-difference images at start time of the type-III bursts for the 2011 February 18 (panel (a)) and 2015 August 24 (panels (b) and (c)) events. Pluses mark the L1 magnetic foot-point. The arrows point to the source flares/jets. The heliographic longitude-latitude grid has 15$^{\circ}$ spacing. (d)--(f) Two-color composite images at start time of the type-III bursts for the 2011 February 18 (panel (d)) and 2015 August 24 (panels (e) and (f)) events. The AIA 304\,{\AA} images correspond to green and the HMI line-of-sight magnetic field (scaled to $\pm$200\,G) to red/black. (\href{https://figshare.com/s/7fcc9303591ddf2746b9}{An animation of this figure is available.}) \label{fig:f3}}
\end{figure*}

Table \ref{tab:tab1} summarizes the characteristics of the examined events. Column 1 gives the particle event start date, column 2 the \isotope[3]{He}-rich period where fluences/abundances were determined. Columns 3 and 4 indicate the average \isotope[3]{He}/\isotope[4]{He} ratio and the average \isotope[3]{He} flux at 10.5\,MeV\,nucleon$^{-1}$, respectively. Note on enormous 1 hr \isotope[3]{He}/\isotope[4]{He} ratios in the early stage of the 2011 February 18 event exceeding a value of 10 at 5.2\,MeV\,nucleon$^{-1}$ (see Figure \ref{fig:f1}(b), left panel). Columns 5 and 6 give the \isotope[3]{He}/\isotope[4]{He} and Fe/O ratio at 0.546\,MeV\,nucleon$^{-1}$, respectively. Column 7 gives the type-III burst start time as observed by Wind/WAVES. Columns 8 and 9 indicate the GOES X-ray flare class and start time, respectively. Column 10 shows the flare location as observed in {\sl SDO}/AIA 304\,{\AA} channel. Columns 11 and 12 give the CME speed and width, respectively, obtained from the SOHO LASCO catalog. 

\begin{figure*}
\epsscale{1.16}
\plotone{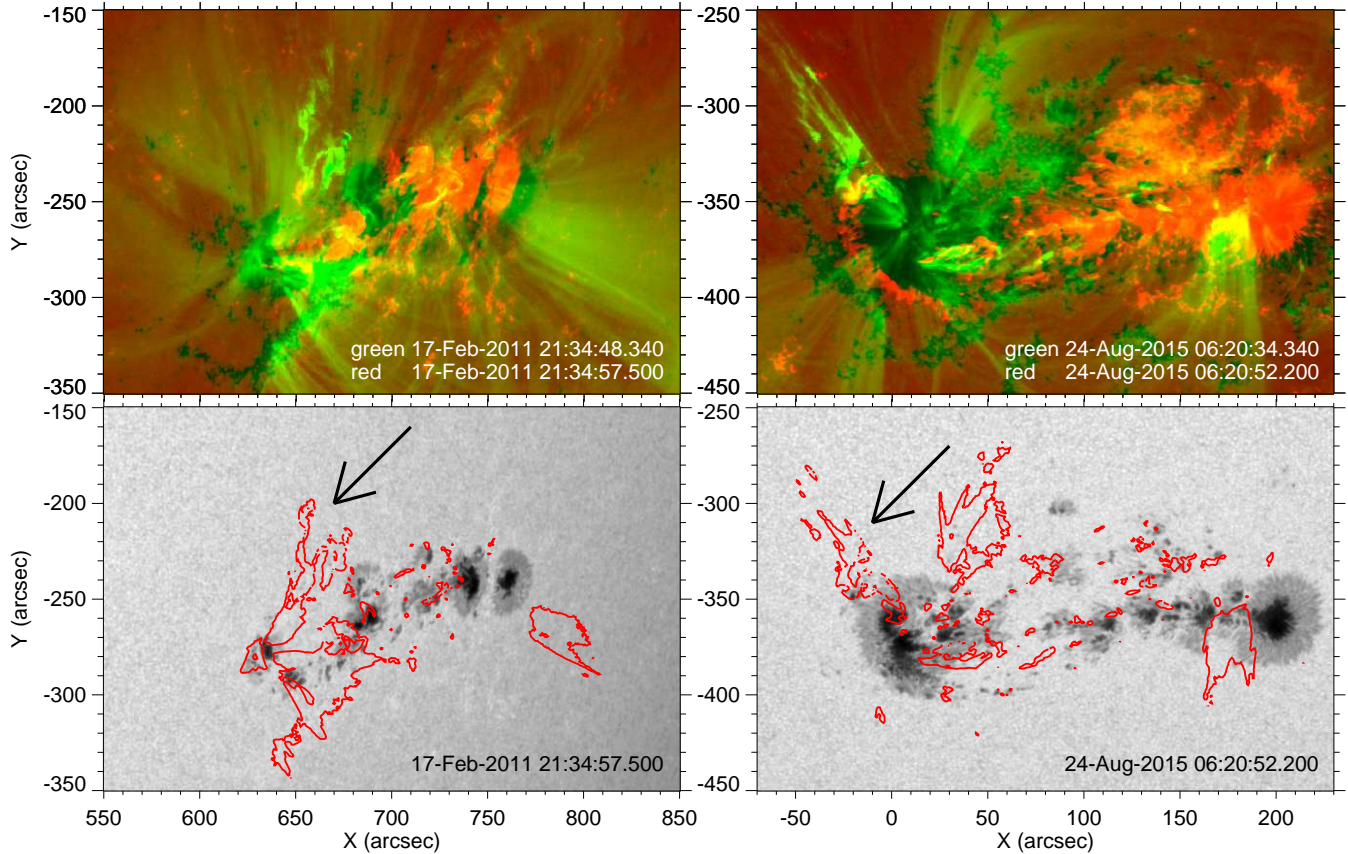}
\caption{Top: Same as Figures \ref{fig:f3}(d), \ref{fig:f3}(e) but here the 171\,{\AA}  image corresponds to green. Bottom: The HMI continuum intensity. Over-plotted contours mark the 95th percentile of intensity in the corresponding 171\,{\AA}  images. The arrows point to the event associated jets. The images are for the 2011 February 18 (left) and 2015 August 24 (right) events. \label{fig:f4}}
\end{figure*}

\subsection{Solar Sources} \label{subsec:ss}

To localize the solar source of the \isotope[3]{He}-rich SEP events, we have examined high-cadence, full-disk  {\sl SDO} EUV images for flaring around the time of the event associated type-III bursts. Figures \ref{fig:f3}(a)--(c) show the base-difference images at 193\,{\AA}, revealing jets at the type-III burst times. Associated animations show the progress of flaring activity for several minutes around the type-III bursts. Note the relatively large longitudinal ($\sim$45$^{\circ}$) separation between the L1 magnetic foot-point and the associated jets in the August 24 event. The foot-point longitude of the Parker spiral connecting L1 was determined from the measured solar wind speed at the type-III burst onset time. In the February 18 event the jet was accompanied by a faint larger-scale front propagating northward; in the August 24 event the jets evolved to a wider eruption.

To have a more detailed view on the event associated jets and the underlying photospheric magnetic field, Figures \ref{fig:f3}(d)--(f) show two-color composite images where the EUV 304\,{\AA} channel corresponds to green and the line-of-sight magnetic field to red (+black). The black and light-red areas indicate regions with strong magnetic fields of negative and positive polarity, respectively. The images are shown at times of the event associated type-III bursts (see the animation for other times). In both events, the EUV jets are rooted at the boundary of a large sunspot at the interface with the minor (positive) polarity field. In the February event, the magnetic fields of opposite polarities were separated by about 15$\arcsec$; in the August event, the fields were closely adjacent. The jet in the February event shows a complex helical structure with unwinding counterclockwise motion. The jets in the August event show a structured spire with no obvious rotation. The jet at the onset of the M5.6 flare in the August 24 event was less clearly visible. The animation shows that type-III burst on August 24 06:16\,UT coincides with the jet at the sunspot boundary and not with C3.2 flare at 06:13\,UT (the first X-ray flux increase in Figure \ref{fig:f1}(a), right panel) that occurred in the more central part of the sunspot along the closed loops. 

Figure \ref{fig:f4} (top panels) shows another two-color composite image combining the EUV 171\,{\AA} channel and the line-of-sight magnetic field. Here coronal loops, connecting the fields of opposite polarities, and peripheral fan structures are clearly seen. Strikingly the jets are not co-located with these large-scale coronal structures which were visible in the sunspots before the events. The jet in the August event appears more complex in 171\,{\AA}, but a rotation (as in the February event) is not observed. Figure \ref{fig:f4} (bottom) shows the corresponding HMI continuum with over-plotted contours of high (95th percentile) level of intensity from the 171\,{\AA} image. The jets in both events appear to be rooted near the sunspot penumbra.

AR 11158, associated with the 2011 February 18 event, emerged in the eastern solar hemisphere on February 11. According to the Solar Event List (\url{ftp://ftp.swpc.noaa.gov/pub/warehouse}), AR 11158 produced 1 B-, 56 C-, 5 M-, and 1 X-class flare during its disk transit. AR 11158 consisted of two major emerging bipoles that showed complex sunspot motion and interaction \citep[e.g.,][]{jia12}. AR 12403, associated with the 2015 August 24 event, also contains two major bipoles. AR 12403 appeared near the eastern limb on August 17. During its disk transit, AR 12403 produced 34 B-, 84 C-, 11 M-class flares. Both ARs show a less common $\beta\gamma\delta$\footnote{$\beta\gamma$ denotes a bipolar sunspot group with no clearly marked line separating spots of opposite polarity; $\delta$ indicates a penumbra enclosing umbrae of opposite polarity} magnetic classification. The $\beta\gamma\delta$ class has been reported in $\sim$4\% of all ARs in 1992--2015 \citep{jae16}.  According to USAF/NOAA Sunspot Group reports (\url{https://www.ngdc.noaa.gov/stp/space-weather/solar-data/solar-features/sunspot-regions/usaf_mwl/}), AR 12403 was among the ten largest ARs in the present solar cycle. The maximum reported area was 1320 millionths of the solar hemisphere. The maximum area of AR 11158 was 670.

\section{Discussion and Summary} \label{sec:dis}

We have examined the solar sources of the two most intense high-energy ($>$10\,MeV\,nucleon$^{-1}$) \isotope[3]{He}-rich SEP events of the current solar cycle. \deleted{Interestingly, $Z$$>$2 ions were below the detection threshold at these energies.}We have found that the solar sources of the \isotope[3]{He}-rich SEPs were structured jets -- in one case with an untwisting motion. A striking feature of the investigated events is their association with jets originating at the boundaries of large and complex sunspots with $\beta\gamma\delta$ magnetic class. The sunspots produced numerous (63 and 129) X-ray flares during their disk transit. The August 24 event was accompanied by four type-III bursts within one hour, suggesting that the extreme \isotope[3]{He} intensity may be related to unresolved multiple ion injections. The February 18 event was associated with only a single type-III burst.

Solar sources of high-energy ($>$10\,MeV\,nucleon$^{-1}$) \isotope[3]{He}-rich SEPs have not been specifically tackled in previous studies. \citet{nit15} have reported that about 1/2 (13 out of 29) of their events have energetic \isotope[3]{He} measured at $>$4.9\,MeV\,nucleon$^{-1}$ where eleven events were associated with jets or ejections and two with coronal waves. One of those events was associated with a helical jet from an active region near a coronal hole \citep{inn16}. Four events in \citet{nit15} that do not show \isotope[3]{He} at SIS high energies were associated with jets from a plage region \citep{che15}. \citet{buc16} have found that a quarter (8 out of 32) of the events in their survey have \isotope[3]{He} measured at $>$7.6\,MeV\,nucleon$^{-1}$ where six events were associated with coronal waves (where a few started as a jet) and two with brightening. No information on the photospheric field has been reported for these events. The only \isotope[3]{He}-rich SEP event associated with sunspot umbral jet did not show an extension of \isotope[3]{He} measurements to the SIS energy range \citep{nit08}.

The earlier works \citep{koc03,tor03} have suggested that high-energies of \isotope[3]{He}-rich SEPs may be due to re-acceleration in coronal shocks. This is unlikely in the events reported here that were accompanied by faint CMEs without a coronal shock signature. An enhancement of suprathermal \isotope[3]{He} in IP shock events has been reported in \citet{des01}. It has been suggested that the \isotope[3]{He} enhancement is due to an acceleration of remnants from prior \isotope[3]{He}-rich SEP events \citep{des01} or a solar particle event is swept up by the shock \citep{tsu02}. Different time-intensity profiles for \isotope[3]{He} and \isotope[4]{He} and a velocity dispersive onset in the February 18 event indicate that the IP shock preceding the event might only marginally affect the \isotope[3]{He} enhancement. An association with jets from the edge of sunspots with complex magnetic configuration may be a distinct feature of high-energy \isotope[3]{He}-rich SEP events. It has been known that a large sunspot area and complex magnetic configuration are important to produce the largest flares \citep[e.g.,][]{sam00}. We plan to address such relation for production of high-energy \isotope[3]{He}-rich SEPs in the forthcoming statistical study.

\acknowledgments
The work of R.B. was supported by DFG grant BU 3115/2-1. The work of M.E.W. was supported under NASA Goddard grant 80NSSC18K0223 to Caltech. Work at JHU/APL was supported by NASA grant NNX17AC05G/125225. R.G.H. acknowledges the financial support of the Spanish MINECO projects ESP2017-88436-R and EPS2015-68266-R. The work by N.V.N. was supported by NSF grant AGS-1259549 and NASA grant 80NSSC18K1126. R.B. and R.G.H. acknowledge the support of the University of Alcal\'a through the Giner de los R\'ios visitor program. We wish to acknowledge Davina Innes for useful discussions and comments.

\end{document}